# The displacement and annihilation of macroscopic regions with hypervortices in ceramic $YBa_2Cu_3O_{7-x}$

A. A. Shablo, V. P. Koverya, and S. I. Bondarenko

*B. I. Verkin Institute of Low-Temperature Physics and Engineering, National Academy of Sciences of Ukraine, pr. Lenina, 47, Kharkov, 61103, Ukraine*

Features of the controllable displacement, annihilation, and transformation of macroscopic regions of a weak frozen magnetic field (FMF) have been experimentally studied in a plate of granulated ceramic $YBa_2Cu_3O_{7-x}$ under the action of the transport current flowing through it. It is shown that the displacement of regions with an FMF lower than the first critical field of the ceramic granules ($H_{c1g}$) is associated with a relatively weak pinning force of the hypervortices that is less than the maximum experimentally attainable Lorenz force. A discussion is presented of the role of the transport current that not only is one of the conditions for the appearance of the Lorenz force but also acts on the internal current structure of the local FMF, which in turn influences its mobility and magnitude. If a local field greater than $H_{c1g}$ is frozen in, the experimentally produced Lorenz force is less than the pinning force of the Abrikosov vortices in the mixed state of the granules, and there is no displacement of the regions with an FMF.

## I. INTRODUCTION

The freezing of a magnetic field in granulated ceramic $YBa_2Cu_3O_{7-x}$ with a low critical current density ($j_{cJ} < 100$ A/cm$^2$) is usually evidence of the distinct separation of a range of small values of the external field, in which the field is lower than the first critical field of the ceramic granules ($H_{c1g}$) and is frozen in the intergranular Josephson medium in the form of hypervortices, and a range of relatively large values (greater than 50 Oe), in which a mixed state is achieved in the granules, and the field is frozen in them mainly in the form of Abrikosov vortices.[1] When the local field is frozen by the technique described earlier[2] and the transport current ($I_{tr}$) is transmitted through the sample, the Lorenz force $\mathbf{F} \propto \mathbf{I}_{tr} \times \mathbf{B}_f$ appears, which acts on the region with a frozen magnetic field (FMF) with magnetic induction $\mathbf{B}_f$ and is capable of displacing it along the sample.[3,4] By fixing the displacement parameters, information on the local values of the pinning force and the viscosity of the motion of the magnetic flux at the freezing point can be most directly obtained for various values of the external magnetic field and for both known freezing regimes. In the first regime [often called the field-cooling (FC) regime], the superconductor is cooled to a temperature below the critical value with a magnetic field switched on, and this is frozen into the superconductor after the external field is switched off. In the second regime [called the zero-field-cooling (ZFC) regime], the cooling is done in a zero or close-to-zero external magnetic field, after which an external field that exceeds the critical field of the beginning of its freezing is briefly turned on. For the subsequent study of the features of the displacement of the local frozen field in a granulated ceramic, it became necessary to answer the following questions: (1) Is it possible to completely displace a region with a frozen field beyond the boundary of a sample made from a granulated ceramic when $I_{tr} > I_{cJ}$? (2) How does the process of annihilation of oppositely directed frozen fields of two regions occur when they approach under the action of the Lorenz force? (3) Can a region with an FMF be displaced over the sample when the transport current is less than critical current of the sample, $I_{cJ}$? (4) Does the reaction of a region with an FMF to the action of the Lorenz force vary as a function of the value of $j_{cJ}$ of the ceramic sample? The goal of this paper was to find answers to these questions.

## II. DESCRIPTION OF THE EXPERIMENTS

The ceramic $YBa_2Cu_3O_{7-x}$ was fabricated by the traditional method of solid-phase reaction of the oxides of its components. The superconducting transition began at a temperature of $T_c = 92$ K, and the critical current density at 77 K was $j_{cJ} = 50-60$ A/cm$^2$ in various samples. Magnetic fields of various directions were frozen into one or several samples (diameter $d = 0.5$ mm) along the $X$ symmetry axis of the ceramic sample in the FC regime as follows: (see Refs. 2–4 for greater detail): A sample in the form of a thin ceramic plate with dimensions $0.5 \times 10 \times 10$ mm could be displaced in the vertical direction by means of a rod coming out of the neck of a cryostat and equipped with a micrometer screw in a narrow slot (about 0.5 mm) between two nonmagnetic modules of a holder cooled to liquid-nitrogen temperature. From one to four pairs of microsolenoids 0.5 mm in diameter and 10 mm long were coaxially mounted inside the modules perpendicular to the plane of the sample; these were electrically connected pairwise in series to create in the gap between them (about 0.5 mm) a magnetic field similar in magnitude and configuration to the field at the center of a long solenoid. When the sample was mounted in the gap between the microsolenoids, the indicated structure of the magnetic-field source in the largest degree made it possible to assume that the resulting external local field was mainly concentrated on a section of the ceramic close to 0.5 mm in diameter. The electrical connection of the individual pairs made it possible to simultaneously impress local fields of the same or the opposite direction on various sections of the sample. Besides this, inside one of the indicated holder modules, a miniature (diameter 0.4 mm, length 4 mm) probe coil with axis perpendicular to the surface of the sample and a gap of about 0.1 mm between the end of the probe coil and the sample surface was mounted at a certain distance from the microsolenoids. The sensitivity of the probe coil to a homogeneous magnetic field was $10^{-4}$ Oe, and to the local field from a source located close to the end of the probe coil was about $10^{-1}$ Oe.

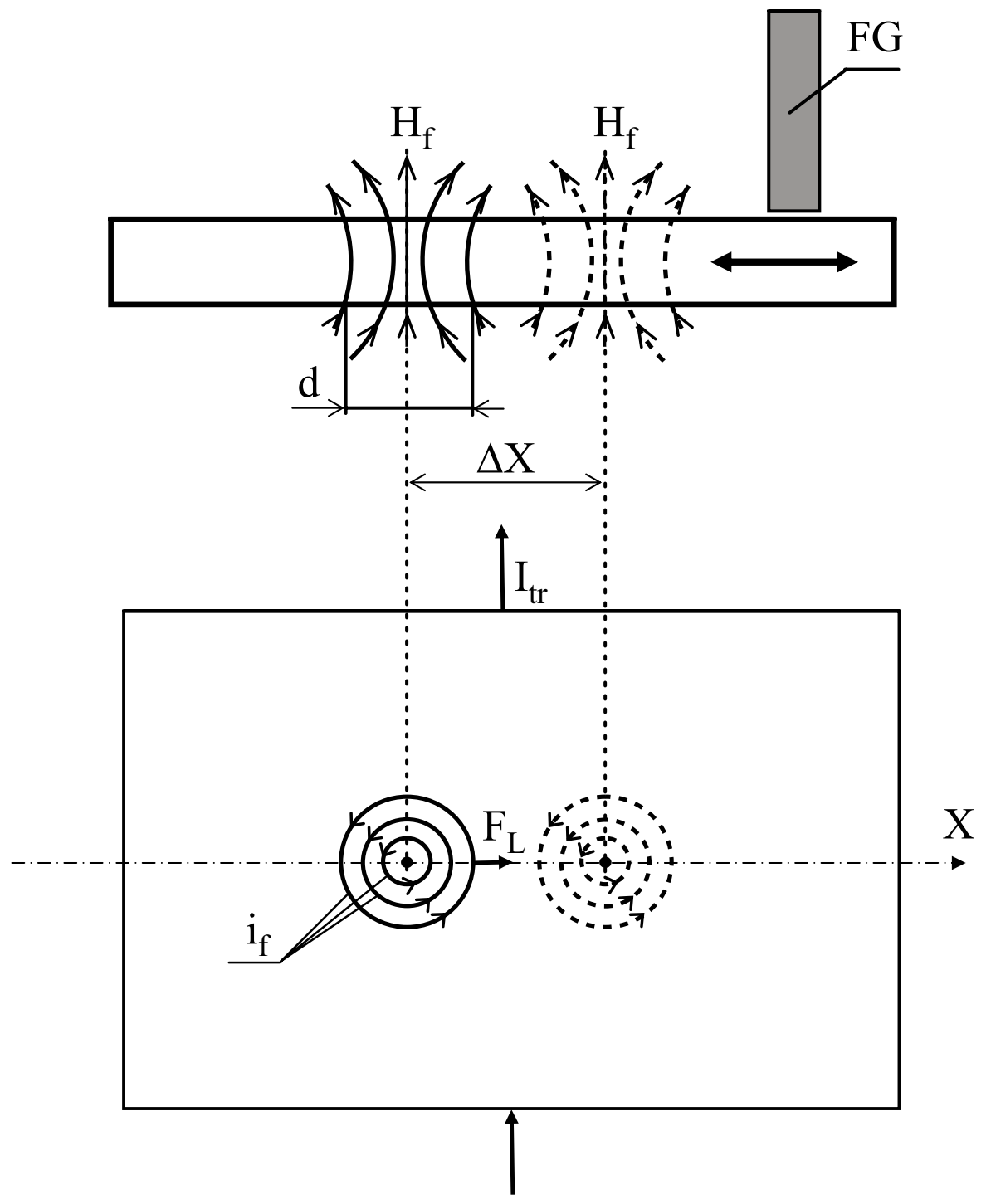

FIG. 1. Layout of a ceramic plate with locally frozen field $H_f$ in a region of diameter $d$, which can be displaced by distance $\Delta X$ under the action of the Lorenz force $F_L$ ($i_f$ is the frozen vortex current, and FG is a probe coil).

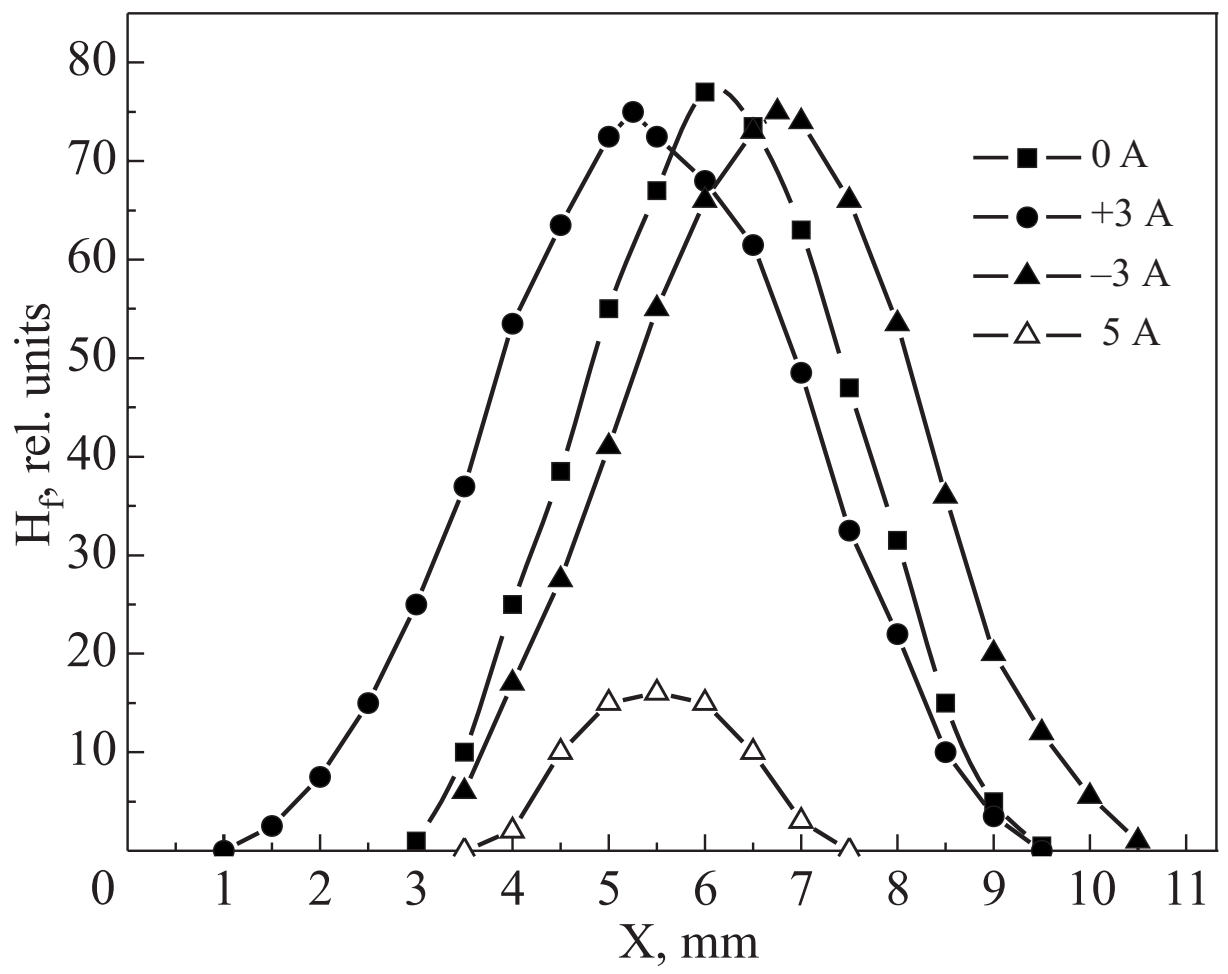

FIG. 2. Curves of the FMF distribution close to the surface of a ceramic sample along the $X$ axis of a ceramic plate (see Fig. 1). The curve corresponding to the initial FMF before current $I_{tr}$ passed through the sample is designated as 0 A, and the other curves were recorded after currents equal to 3 A, −3 A, and 5 A passed through the sample. The last curve corresponds to the residual part of the FMF, which was displaced virtually not at all. The fact that the scattering field is recorded along the entire length of the sample and not only in the freezing region with a diameter of about 0.5 mm is explained by the spatial elongation of the scattering field of the source of the FMF and by the elongated probe-coil detector (4 mm) in a direction perpendicular to the plane of the sample.

The displacement of a region with an FMF along the sample under the action of the transport current was tracked by monitoring the position of the maximum (or minimum) of the spatial distribution of the vertical component of the FMF ($H_f$) close to the surface of the sample; the maximum (or minimum) was assumed to coincide with the center of the region. The distribution of $H_f$ above the surface of the sample along its $X$ axis was measured by means of a probe coil, relative to which the sample was displaced. The layout of a sample with one region that contains a frozen field before and after current $I_{tr}$ is transmitted is shown in Fig. 1. A Permalloy magnetic shield kept the magnetic field stable at $10^{-3}$ Oe in the region of the sample. The transport current through the sample could be adjusted in the range 0−5 A and, to avoid the possibility of overheating the sample, was transmitted through it briefly (for about 1 sec). The ceramic was at liquid-nitrogen temperature (77 K), while the external local field perpendicular to the surface of the sample was 25 Oe, and this is less than the first critical field of the ceramic granules at the given temperature (50 Oe). The critical current density of the sample was varied by decreasing it to $0.5 \times 10^{-3}$ A/cm$^2$ by annealing the sample in air at a temperature of 400 °C for two hours.

### III. EXPERIMENTAL RESULTS AND DISCUSSION

To determine the possibility of displacing the local FMF beyond the limits of the sample, a single section with a frozen field was formed in the central part of the sample by the technique described above. The distribution along the $X$ axis of the sample of the component of the FMF normal to the plane of the sample with an epicenter at a point 6 mm along the horizontal axis is shown in Fig. 2 (the 0-A curve). The reaction of the FMF to the action of a transport current that slightly exceeds the critical current ($I_{tr}$=3 A, $I_{cJ}$=2.8 A) was first checked. The maximum of the FMF distribution was displaced to the left by 0.9 mm (the +3-A curve). The sample was next heated to a temperature of $T > T_c$, and a section again formed with the same FMF and an epicenter at the 6-mm point. A transport current was then supplied, with the same magnitude but in the opposite direction. In accordance with the change of direction of the Lorenz force, the maximum of the FMF distribution moved to the right by the same 0.9 mm, with the new position of the epicenter at the 6.9-mm point. These experiments demonstrated that the process of displacing a region with an FMF is reproducible and made it possible to estimate the spatial scale of the possible displacement as a function of current $I_{tr}$. Starting from the assumption that the displacement is proportional to $I_{tr}$ and to the remaining distance to the edge of the sample, the next $I_{tr}$ pulse was 5 A. A survey of the section of the sample to the right of the 6.9-mm point using the probe coil showed that the FMF was virtually completely absent. According to our assumption, this is because the greater part of the FMF (80%) was displaced from the 6.9-mm point of the epicenter beyond the edge of the sample (the 10.5-mm point on the horizontal axis of Fig. 2). It was simultaneously detected that, after a 5-A pulse was supplied, a residual frozen field remained near the original position of the region with the FMF (near the 6-mm point), with the epicenter of the distribution close to the initial 6-mm point (the 5-A curve—open triangles); this is an argument in favor of the assumption made above that a movable part of the FMF exists. In what follows, it is proposed to study the stages of the FMF displacement for several values of the current $I_{tr}$ with a smaller step between them. Thus, if a 3-A pulse displaced the region with an FMF by 0.9 mm, a pulse that is stronger by a factor of 1.7 can displace the movable part of it by a distance at least a factor of 4 greater. The nonlinear dependence of the displacement on $I_{tr}$ can be associated with the suppression (increasing as $I_{tr}$ increases) of the superconductivity of part of the weakest intergranular bonds of the ceramic and the corresponding decrease of the viscosity of the motion of hypervortices through the ceramic "weakened" by the current. The maintenance of a smaller, virtually immovable part of the FMF after the 5-A pulse can be explained by the fact that, in the granulated ceramic $YBa_2Cu_3O_{7-x}$, even a weak magnetic field can be frozen in the FC regime not only in the intergranular medium, but also in the form of current vortices in the location of ceramic granules in which the continuity has broken down and which possess a large pinning force by comparison with the indicated medium.[5]

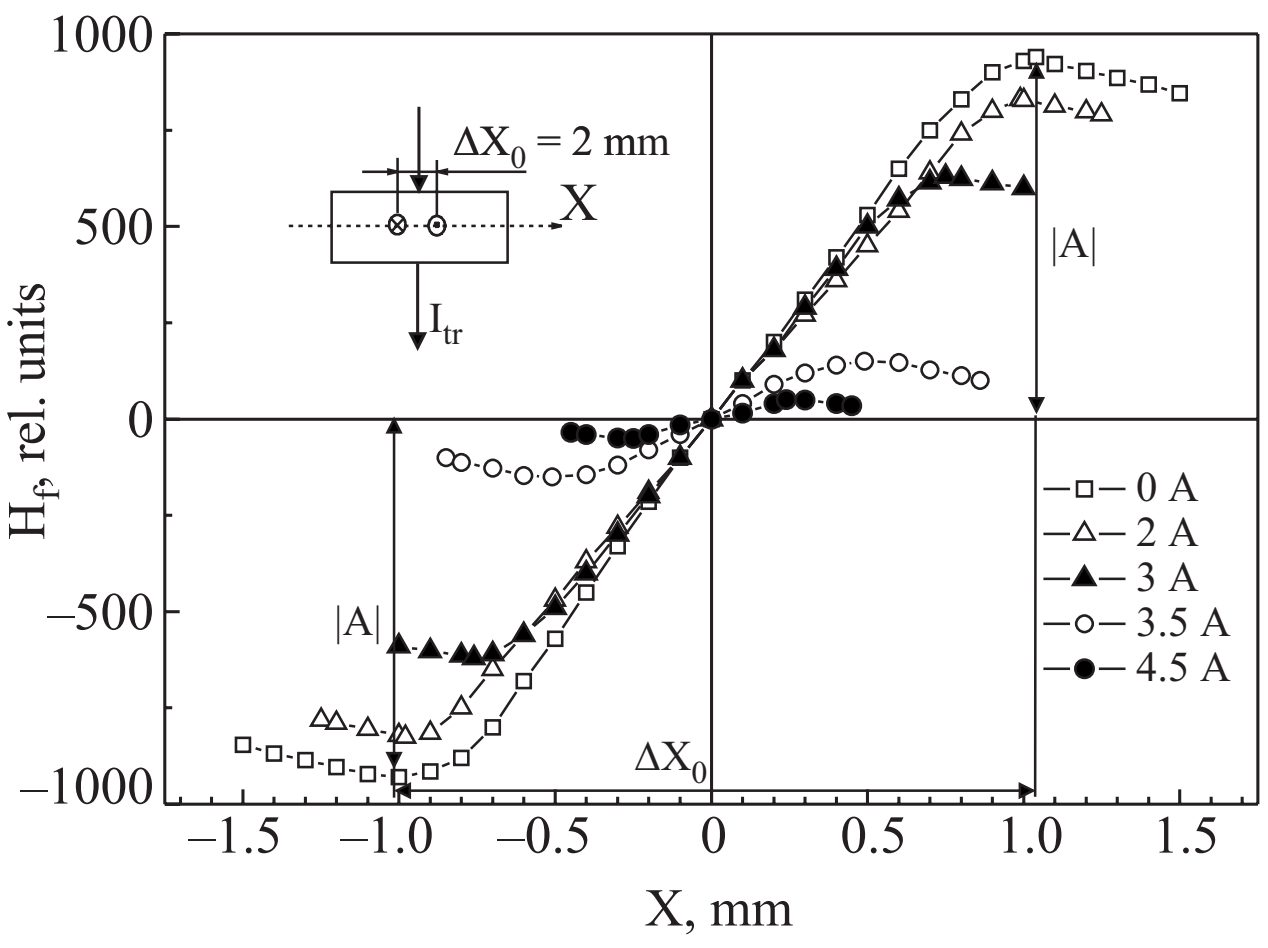

FIG. 3. Variation of the FMF distribution $H_f$ above two regions with hypervortices of different direction (antihypervortices) along the $X$ axis of a ceramic plate (shown in the inset) after a transport current ($I_{tr}$) of magnitude 2, 3, 3.5, 4.5 A briefly passes through the plate ($\Delta X_0$=2 mm is the initial distance between the extrema).

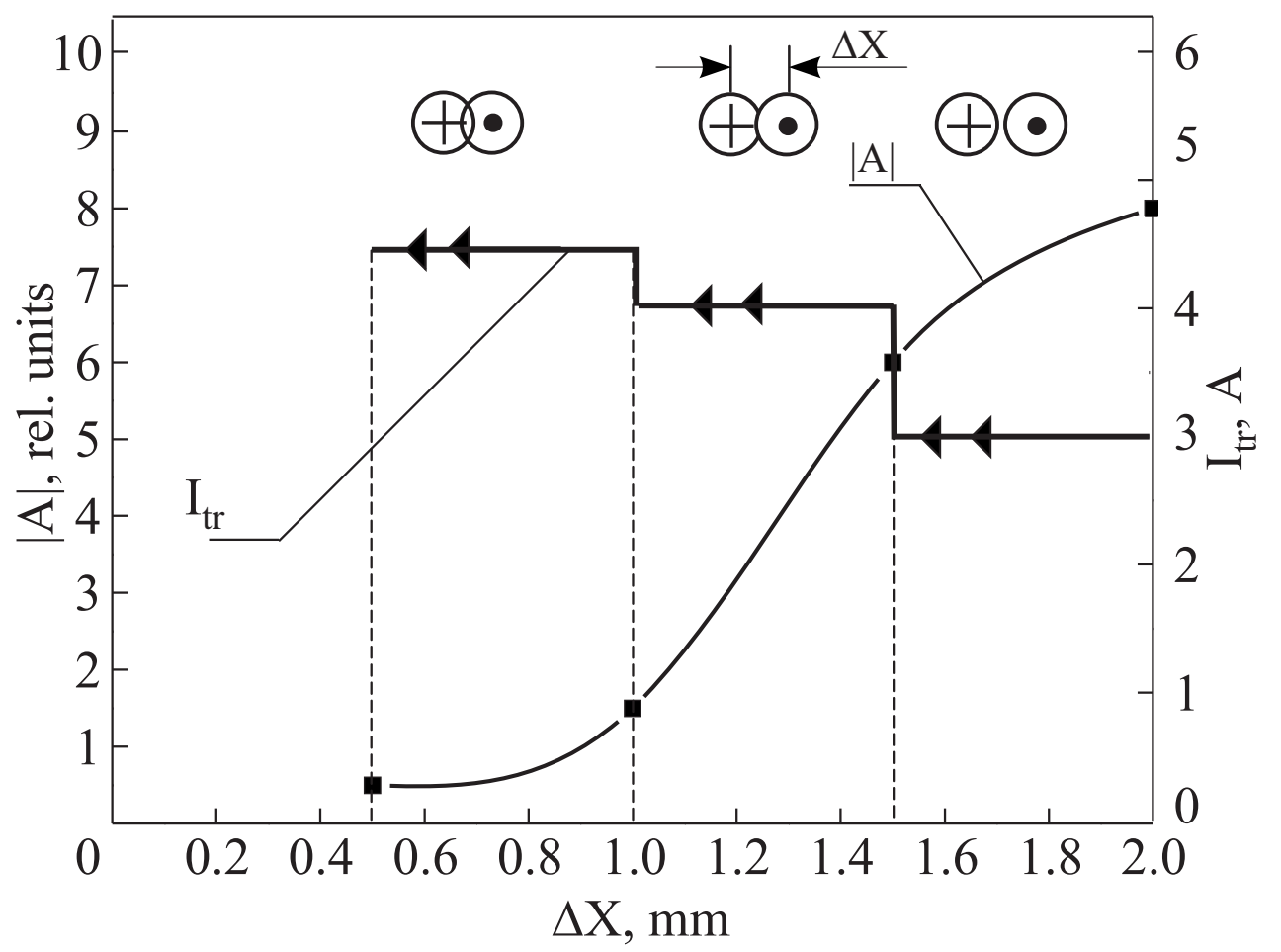

FIG. 4. Dependences of the transport current $I_{tr}$ needed to bring two antihypervortices together by 0.5 mm and the extrema of the magnetic field distribution ($|A|$) of the antihypervortices as a function of the distance $\Delta X$ between them. The upper part of the figure shows diagrams of antihypervortices approaching each other as $I_{tr}$ increases.

When fields that are identical in magnitude but opposite in direction are frozen in two regions (0.5 mm in diameter) of a sample and a transport current is briefly transmitted in a direction perpendicular to the line that connects them, these regions during the indicated time are displaced opposite to each other in complete accord with the direction of the Lorenz force for each of the regions with an FMF. The transmission of several successive current pulses gradually reduces the distance ($\Delta X$) between the extrema of the FMF distribution. At the same time that the distance decreases, there is a reduction of the absolute magnitude $|A|$ of the maximum and minimum of the FMF distribution of the indicated regions (Fig. 3). It can be seen that, if the distance between the regions with FMFs is reduced by a factor of 4, the recorded frozen field at the points of the extrema decreases by a factor of 10. The sharp decrease of the field of each of the regions as they approach each other and taking into account the increasing values of $I_{tr}$ can be explained by the simultaneous action of two factors. First, as the regions with FMFs that are opposite in direction approach each other, they begin to partially compensate each other. Second, the FMF may decrease because of the partial and increasing suppression of the superconductivity of the weak intergranular bonds of the ceramic by the increasing transport current $I_{tr}$ itself.

As a result, higher and higher currents $I_{tr}$ are required to bring regions with FMF together by the same distance (Fig. 4).

The detected feature that the transport current needed for displacing the antihypervortices must be increased can be treated as a certain additional retardation of the hypervortices that appears as they approach each other.

A third feature of the displacement of the local FMF under the action of the Lorenz force was the fact that a region with a frozen magnetic flux located in the central part of the sample can be displaced starting with a transport current equal to $0.7I_{cJ}$ (about 2 A). Earlier, we used a probe coil with coarser spatial resolution to observe[3] the displacement of a region with an FMF only with $I_{tr}>I_{cJ}$ ($I_{cJ}$ is the critical current of the sample, in this case equal to 2.8 A). The resulting value of the transport current thus makes it possible to more accurately determine the pinning force of a region with a hypervortex in the given ceramic. Moreover, Fig. 3 shows the displacement of two regions with oppositely directed FMFs from their position under the action of $I_{tr}$ =2 A when the distance between them is 2 mm. Taking into account that the interaction of two such regions is weak, it is possible to consider the displacement of each of the hypervortices to be independent. It can be seen from Fig. 3 that the displacement of each of the regions is no more than 0.05 mm. On the other hand, a pulse of $I_{tr}$=3 A, somewhat greater than the critical value, as shown above, causes a dis-placement by 0.9 mm. Thus, increasing $I_{tr}$ by a factor of 1.5 in the region in which the sample goes from the superconducting to the resistive state increases the displacement by a factor of 18. If it is taken into consideration that the distribution of the superconducting transport current over the cross section of a sample made from a granulated ceramic changes very little when going from the superconducting to the resistive state,[6] it can be concluded that the resistance of the hypervortex motion sharply decreases as one goes from the superconducting to the resistive state. When the current is less than the critical value, the displacement of the hypervortex is hindered and, like magnetic-flux creep in a classical type-2 superconductor, at the beginning of resistive transition and with a current greater than the critical value, the hypervortex displacement can be compared to the flow process of the flux with some viscosity that decreases as the transport current increases. The reason that the transport current has such an effect is that it can suppress the superconductivity of part of the weak bonds of the Josephson medium of the ceramic and that this suppression increases as the current increases.

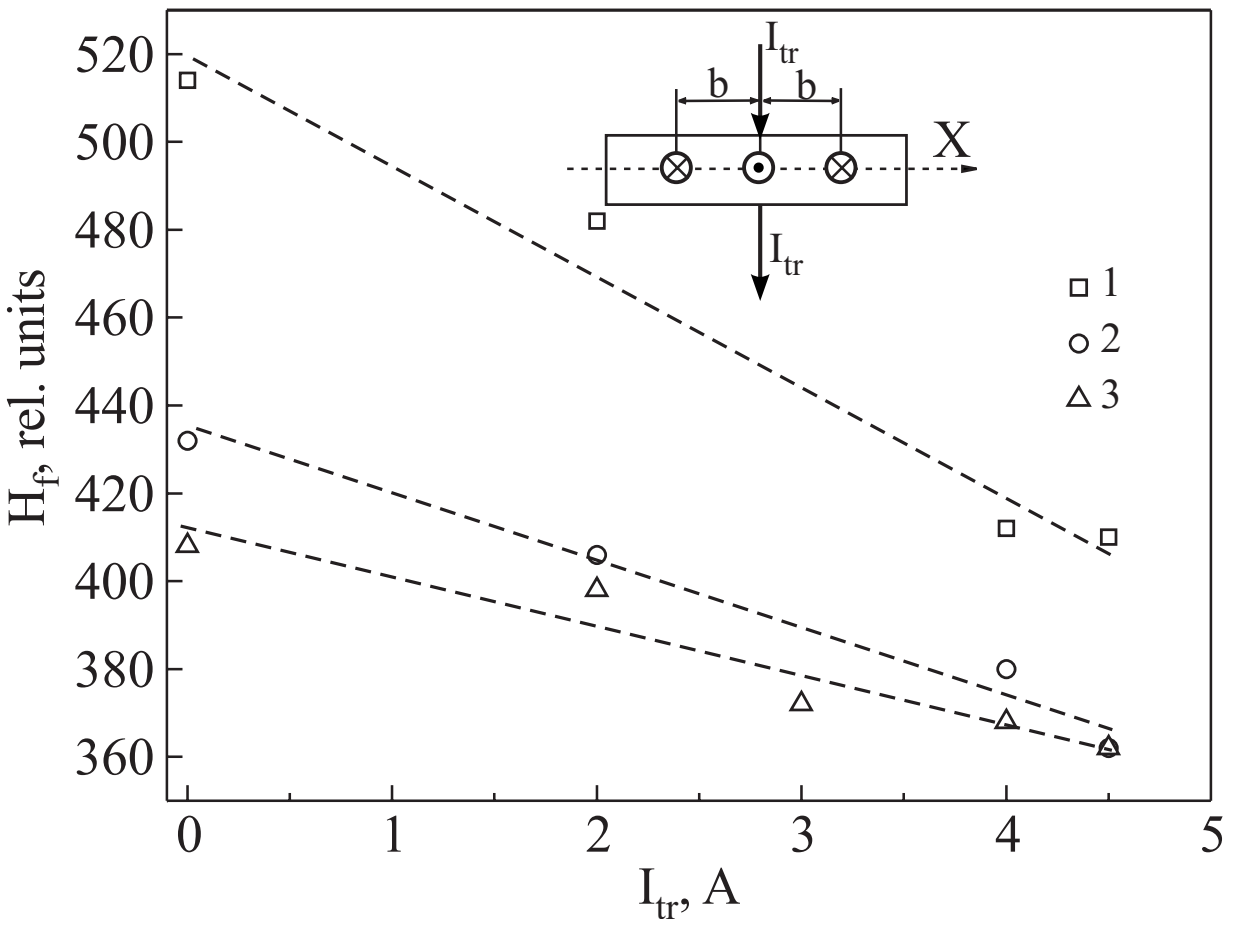

FIG. 5. The absolute value of an FMF above the epicenter of three regions of a ceramic (1, 2, 3) with a critical current density of $0.5 \times 10^{-3}$ A, located along the $X$ axis of the sample at a distance from each other of $b=2$ mm, vs the short-term transport current $I_{tr}$ imposed and directed perpendicular to the $X$ axis through the sample. The differences in the dependences are associated with inhomogeneity of the ceramic over the area of the sample.

To answer the fourth question, the action of the Lorenz force on a region with an FMF in samples with critical current density $j_c \approx 50$ A/cm$^2$ (such regions were used when solving the first three questions) was compared with that in samples with $j_c=0.5\times10^{-3}$ A/cm$^2$, having identical values of the FMF and identical sizes. Although displacement of the region with FMF was observed for the former samples starting from $I_{tr}=0.7I_{cJ}$ and going to $I_{tr}\approx 1.7I_{cJ}$ (which corresponds to a current of 5 A through the sample and a developed maximum Lorenz force of about $5\times10^{-7}$ N), there was no displacement in the range $I_{tr}=0-4.5$ A for the latter samples. Transmitting a transport current in this range of values produced only a small decrease (15–20%) of the FMF of immobile regions (Fig. 5). The indicated features are caused by a substantial degradation of the superconducting properties not only of the intergranular contacts but also of the granules themselves. In particular, measurements of the critical temperature of such samples showed that it decreased from 92 to 83 K. Accordingly, the first critical field of the ceramic granules was reduced. As a result, it can be assumed that a weak external field of 25 Oe exceeds $H_{c1g}$ and is mainly frozen in the ceramic granules in the form of Abrikosov vortices and current vortices of the inhomogeneities of the granules,[5] whose pinning forces are significantly greater than the pinning of the Josephson hypervortices in the intergranular space. In this case, the part of the FMF frozen in the given Josephson medium can be reduced by the gradual breakdown of the remaining weak bonds by the increasing transport current.

## IV. CONCLUSION

The displacement of the hypervortices in the granulated ceramic $YBa_2Cu_3O_{7-x}$ made it possible to detect the following features of this substance:

- The resistance to the displacement of a region with an FMF in the form of a hypervortex sharply decreases as the ceramic sample goes from the superconducting to the resistive state.
- As the transport current through a sample in the resistive state increases and the distance decreases between two regions with equal but oppositely directed FMFs in the form of antihypervortices, there is a nonproportionally strong decrease of the value of the FMF of the two regions. As a result, as the antihypervortices approach and annihilate each other, a higher and higher transport current is required for them to be displaced by the same distance.

When an external local magnetic field that exceeds the first critical field of the granules, $H_{c1g}$, acts on the ceramic, it is mainly frozen into the ceramic granules (in the form of Abrikosov vortices and vortices of macroscopic inhomogeneities of the granules). The region with an FMF is not displaced along the sample in this case, since the Lorenz force produced in experiment, although it is sufficient for displacing the hypervortices, is less than the pinning force of the indicated vortices. There is simultaneously some decrease of the value of the FMF under the action of the transport current.

It follows from an analysis of these results that the transport current through the sample not only determines, along with the value of the FMF, the Lorenz force, but can also directly affect the structure of the currents that maintain the FMF. Most importantly, this affects transport currents that exceed the critical value for the sample. By suppressing the superconductivity of even part of the weak bonds of a Josephson medium with an FMF, the transport current can not only reduce the local FMF, but can also alter the conditions for its displacement along the sample under the action of the Lorenz force. The first type of action can explain the decrease of an FMF that exceeds $H_{c1g}$, because part of it consists of hypervortices suppressed by the transport current. A similar action makes it possible to explain the anomaly in the decrease of the FMF of antihypervortices as they approach. The second type of action is directly related to the sharp increase of the mobility of the FMF when there is a transition from the superconducting state of the sample to the resistive state and can explain the assumed emergence of the FMF from the sample under the action of a 5-A current. In subse-quent studies of the controlled displacement of regions with a given FMF in superconductors, all this will cause the study of how the transport current affects the internal structure of a local FMF to be moved to the forefront.

[1]Kh. R. Rostami, V. V. Mantorov, and V. I. Omel'chenko, Fiz. Nizk. Temp. **22**, 736 (1996) [Low Temp. Phys. **22**, 565 (1996)].
[2]S. I. Bondarenko and A. A. Shablo, in *Collection of Extended Abstracts of the First International Conference on Fundamental Problems of High-Temperature Superconductivity, 18–22 October 2004*, Zvenigorod, Russia, p. 264.
[3]S. I. Bondarenko, A. A. Shablo, and V. P. Koverya, Fiz. Nizk. Temp. **32**, 825 (2006) [Low Temp. Phys. **32**, 628 (2006)].
[4]A. A. Shablo, V. P. Koverya, and S. I. Bondarenko, in *Collection of Extended Abstracts of the Third International Conference on Fundamental Problems of High-Temperature Superconductivity, 13–17 October 2008*, Zvenigorod, Russia, p. 110.
[5]G. Deutscher and K. A. Muller, Phys. Rev. Lett. **59**, 1745 (1987).
[6]V. P. Koverya and S. I. Bondarenko, Vopr. At. Nauki Tekh., Ser.: Tekh. Fiz. Eksp. **1**, 52 (2008).